\documentclass[aps,prl,twocolumn,superscriptaddress,amsmath,amssymb]{revtex4-2}

\usepackage[T1]{fontenc}
\usepackage[utf8]{inputenc}
\usepackage{lmodern}
\usepackage{graphicx}
\usepackage{xcolor}
\usepackage{bm}
\usepackage{siunitx}
\usepackage{hyperref}
\usepackage{placeins}
\usepackage{orcidlink}
\usepackage{float}
\hypersetup{colorlinks=true,citecolor=blue,linkcolor=blue,urlcolor=blue}
\begin{document}

\title{Hot-Layer Ignition in High-Gain Inertial Confinement Fusion}

\author{M.Lourmande\orcidlink{0009-0004-2665-2469}}
\email{matias.lourmande2@cea.fr}
\author{B.Canaud\orcidlink{0009-0000-4342-7635}}
\email{benoit.canaud@cea.fr}
\affiliation{CEA, DAM, DIF, F-91297 Arpajon, France}
\affiliation{Universit\' e Paris-Saclay, CEA, LMCE, F-91680 Bruy\`eres-le-Ch\^atel, France}

\date{\today}

\begin{abstract}
Ignition of the shocked deuterium tritium shell is identified during the thermonuclear combustion of very high-gain targets in inertial confinement fusion (ICF). This mechanism is due to the local energy deposition of the fusion alphas created in this shocked shell, independently of hot-spot ignition, and generates sufficient energy to fully recompress the hot spot. A new analytical criterion for the ignition of the hot layer is presented. A simple three zone model, based on the thick layer hot-spot model, reproduces this mechanism, which breaks down into successive stages: ignition of the hot spot, compression of the shocked shell, ignition of the shocked shell, and subsequent recompression of the hot spot. 1D numerical simulations highlight the onset of this hot layer ignition thereby satisfying the analytical ignition criterion. It is shown that this mechanism amplifies the energy released by the combustion of the target by approximately $50\%$.
\end{abstract}

\maketitle

\begin{figure}[h!]
  \centering
  \includegraphics[width=\linewidth]{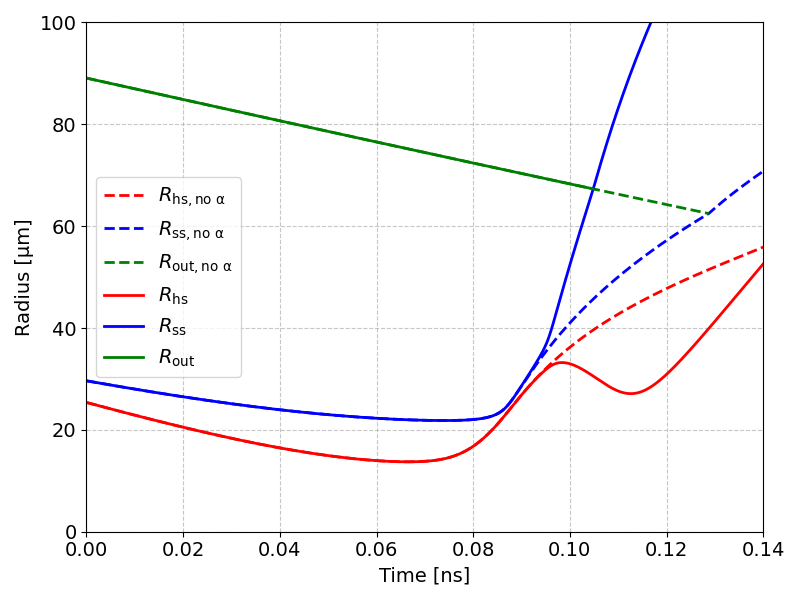}\\
 \includegraphics[width=\linewidth]{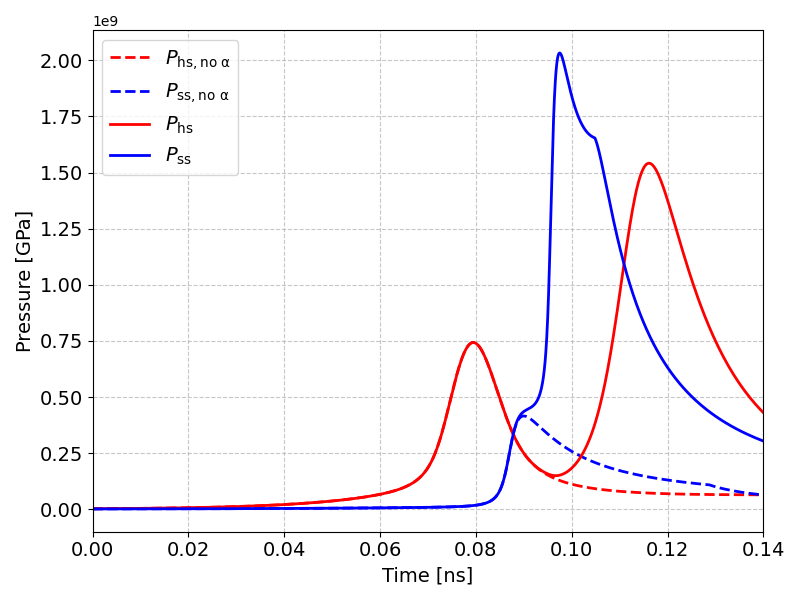}%
  \caption{Target trajectories (top) of hot spot (red line), return shock front (blue line) and free fall shell outer side (green line) given by the model. Dashed lines represent the same trajectories in the case without redeposition of alpha energies in the shocked shell. Time evolution of the pressure (bottom) in the hot spot (red line), the shocked shell and the free falling layer.}
  \label{fig:model}
\end{figure}

\begin{figure}[h!]
  \centering
  \includegraphics[width=\linewidth]{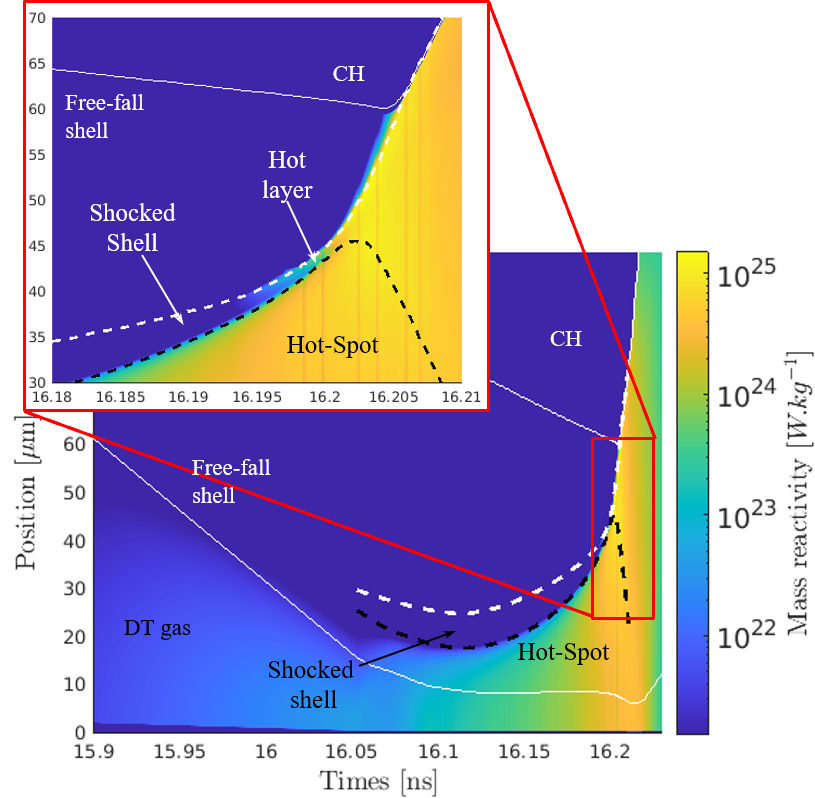}%
  \caption{Spatiotemporal evolution of the fusion power per unit mass for the target design described in \cite{2024_Temporal_epjplus}. The small picture is a zoom in on the the igniting shocked shell. The interface of the hot spot and the shocked shell are highlighted in black and white dotted lines respectively. The hot spot interface is defined as $\text{max}(-\nabla(\rho)\nabla(T_e))$ and the shocked shell interface is defined as $\text{max}(-\nabla v)$.}
  \label{simul}
\end{figure}

\begin{figure}[h!]
  \centering
  \includegraphics[width=\linewidth]{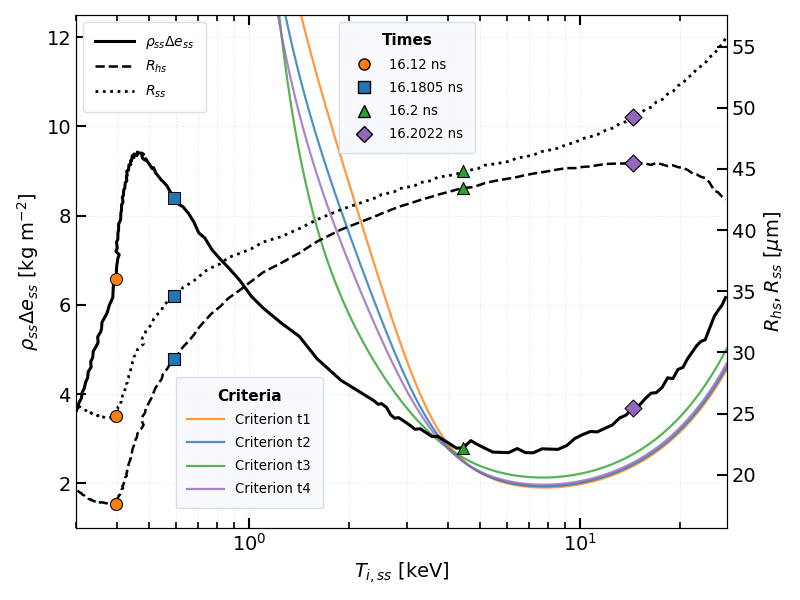}%
  \caption{Comparison of the ignition criterion from Eq.\ref{couche_critere} to the time evolution of $\rho_{ss}\Delta e_{ss}$ versus $T_{i,ss}$ extracted from MULTI-IFE simulations (red plain line). The hot-spot radius (red dashed line) is also plotted versus $T_{i,ss}$}
  \label{ign_crit}
\end{figure}

Hot-spot ignition is the standard route to high thermonuclear gain in inertial confinement fusion (ICF)~\cite{1972_nuckolls_nature,1995_lindl_PP,Atzeni2004}. During deceleration, the kinetic energy of an imploding cryogenic shell is converted into the internal energy of a low-density central hot spot. If alpha particle heating overcomes hydrodynamic, conductive, and radiative losses, the hot spot enters a self-heating regime and burn can extend into the surrounding dense deuterium tritium (DT) fuel. In this conventional description, the hot spot initiates the burn, whereas the surrounding shell is primarily regarded as a confined fuel reservoir.\\
The dense fuel is nevertheless thermodynamically structured during stagnation. An outward propagating shock separates an inner shocked shell ($ss$) from the remaining freely falling shell ($ff$)~\cite{2004_Garnier_PoP,2005_Sanz_PoP,2015_Betti_PRL,Betti_2016,2018_Christopherson_PoP,2020_Christopherson_PoP}. The shocked layer can simultaneously possess a density much larger than that of the hot spot ($hs$) and a temperature high enough for appreciable DT reactivity. Because the volumetric fusion rate scales as $n_D n_T\langle\sigma v\rangle$, reactions generated throughout this finite volume can produce a substantial local alpha heating source.\\
Recently \cite{2026_Lourmande_PoP}, whilst studying a solid fuel at room temperature ICF target design – consisting of carbon hydride (CDT) – we observed ignition occurring in this shocked layer, separated from the hot spot by a temperature gap, which released sufficient energy locally – the mean free path of alpha particles in this layer being virtually zero given its particularly high density – to recompress the hot spot. The question of the existence of this hot igniting layer therefore naturally arose for pure deuterium tritium (DT) fuels.\\
It is generally accepted \cite{Betti_2016,2018_Christopherson_PoP,2020_Christopherson_PoP,2026_Goncharov_PoP} that, when the hot spot ignites, a thermonuclear combustion wave propagates through the shocked layer via thermal conduction, with the deposition of fusion-produced alpha particles directly at the edge of the hot spot contributing to its propagation. However, the increase in pressure within the hot spot compresses the shocked region, thereby simultaneously increasing its density and ionic temperature, whilst the shock propagates outwards into the imploding DT layer. In this shocked shell, fusion reactions can take place, producing alpha particles which locally redeposit their energy into the layer, given the conditions of density ($\rho \sim 1000 $g/cm$^3$) and electron temperature ($T_e < 1 keV$). When the thermonuclear power redeposited by these alpha particles becomes sufficiently high, the shocked layer may potentially ignite, causing its internal pressure to rise sharply until it overcomes the hot-spot pressure, leading to the recompression of the latter, if this pressure reversal lasts long enough. This recompression has already been observed in previous numerical studies \cite{2004_gauthier_pre,2025_Lan_PoP,2026_Lan_PPCF}, although the existence of this hot layer ignition has never been evidenced.\\
The thermodynamic conditions for ignition of the shocked shell of average density $\rho_{ss}$ and thickness $\Delta e_{ss}$, at ionic $T_{i,ss}$ and electronic  $T_{e,ss}$ temperatures, are derived directly, in the same way as those determined for the hot spot, from a power balance for this layer, which is the difference between the losses due to heat conduction and radiation and source terms due to the fusion power redeposited in the shocked shell. This ignition criterion concerns areal density of the shocked shell $\rho_{ss} \Delta e_{ss}$ and is expressed as:
\begin{align}
&    \rho_{ss} \Delta e_{ss}  > \sqrt{\dfrac{2\left(A_{ss}^2+A_{ss}+\dfrac{1}{2}\right) }{\left(A_{ss}^2+A_{ss}+\dfrac{1}{3}\right)\pi}}\nonumber\\
    &\times \sqrt{\dfrac{\kappa_{\text{cond,ss}}\eta_{e,ss}(k_B T_{e,ss})^{7/2} m_{DT}^2}{\left[(\mathcal{P}_{\alpha dep} (T_{i,ss}) - 4\eta_{rad} C_r\sqrt{k_B T_{e,ss}}\right]}},
    \label{couche_critere}
\end{align}
where $\mathcal{P}_{\alpha dep}(T_{i,ss})$ is the fusion power redeposited in the shocked shell by $\alpha$ particles produced by this region, assuming that the neutrons escape from the target without depositing any significant amount of energy. $\kappa_{\text{cond,ss}}$ is the classical Spitzer conductivity\cite{1953_spitzer_PR}, $\eta_{e,ss}$ is the ratio of the gradient length of the electronic temperature profile to shell thickness, $\eta_{rad}$ is a coefficient that takes account of the reabsorption of radiation within the shell,  and $A_{ss}$ is the aspect ratio of the shell, defined as its inner radius over its thickness. The heat flux originating from the hot spot and directed towards the hot layer is not considered a source term, as the mass density at the edge of the hot spot peaks at such a high value that it thermally isolates the layer from the hot spot, and the heat flux is entirely consumed by the ablation process occurring from the hot spot outwards.\\
When this ignition criterion is exceeded in the shocked shell, the pressure increases considerably over time, in accordance with the equation describing this shell: 
\begin{equation}
\frac{d p_{ss}}{dt}
=(\gamma-1)\mathcal{S}_{ss}
-\gamma p_{ss}\frac{\dot V_{ss}}{V_{ss}},
\label{eq:pressure_general}
\end{equation}
which is an extension of the model developed in Ref.\cite{Betti_2016,2018_Christopherson_PoP,2020_Christopherson_PoP}, but to which a source term $\mathcal{S}_{ss}$ has, now, been added to account for the locally energy deposition of alpha particles in the shocked shell and also radiative losses. The approximation, which considers the local deposition of alpha-particle energy, tends to overestimate self-heating when the alpha stopping length is not small compared with the shocked-shell thickness.\\
In this model and following Ref.\cite{Betti_2016,2018_Christopherson_PoP,2020_Christopherson_PoP}, the hot-spot boundary obeys
\begin{align}
M_{\rm eff}\dot U_{\rm hs}&=4\pi R_{\rm hs}^{2}\left(p_{\rm hs}-p_{\rm ss}\right),\label{eq:radius}\\
M_{\rm eff}&=M_{\rm hs}+\alpha_{\rm eff}M_{\rm ss},\label{eq:M_int}\\
\dot R_{hs} &= U_{hs} + \frac{\dot{m}}{4\pi R_{hs}^2 \rho_{ss}}, \label{eq:acc}
\end{align}
where $\alpha_{\rm eff}$ represents the fraction of shocked-shell mass dynamically coupled to the interface and $\tfrac{\dot{m}}{4\pi R_{hs}^2\rho_{ss}}$ represent the ablation velocity due, in major, to thermal conductivity. Equation~\eqref{eq:radius} makes the causal condition explicit: the interface accelerates inward once $p_{\rm ss}>p_{\rm hs}$. This helps to slow the spread of the hot spot. If this negative acceleration is sufficient, it can reverse the direction in which the hotspot spreads, this will tend toward recompression of the hot spot towards the centre of the target, as the layer explodes, amplifying the combustion of the hot spot, as shown in Ref.\cite{2004_gauthier_pre,2026_Lan_PPCF}. The evolution of the shock front radius is directly given by the Rankine–Hugoniot equations, whilst the outer radius of the free falling layer is not assumed to be affected by the increase in pressure within that layer.\\
The figure \ref{fig:model} illustrates this mechanism over time. To do this, we have considered the target described in Ref.\cite{2024_Temporal_epjplus}, at the initial condition taken at the point of maximum kinetic energy during implosion, in the form of three distinct layers: a central hot spot, a shocked shell overlaid by a free falling imploding layer. 
During the first part of the deceleration, the radius of the hot spot decreases until it comes to the stagnation, and the shock front propagates through the free falling layer. At this moment, the central part – the hot spot – ignites and strongly repels the shocked part, which becomes increasingly compressed. When the fusion reactions produce and redeposit sufficient energy in the shocked shell, the pressure inside it overcomes the hot-spot pressure (see Fig.\ref{fig:model}-bottom), and the hot-spot radius is seen to decrease once more. The same graph shows the trajectories of the shock front and the radius of the hot spot (dashed lines), without taking into account the redeposition of alpha particle energy in the shocked shell. It can be seen that, in that case, the hot spot is no longer recompressed and all the interfaces are expanding outwards.\\
We observe that a pressure inversion also occurs during the deposition of alpha particles within the shocked layer. However, this does not lead to the recompression of the hot spot. To achieve recompression, the energy balance of the hot layer must be positive.\\
It is interesting to examine the power balance of each region of the target in more detail. Indeed, the net power released by the hot spot, whilst positive at the moment of ignition, decreases and becomes negative during its expansion because the propagation of the thermonuclear combustion wave is insufficient to compensate for the losses due to the expansion of the hot spot. When the hot layer ignites, there is a renewed increase in the power released by the hot spot, which once again becomes positive. Without alpha energy redeposition in the shocked shell, The total power is positive and reaches a maximum at time of hot-spot ignition, then becomes negative throughout the target’s decompression. In terms of energy balance, the recompression of the target, accompanied by the explosion of the shocked shell, can doubles the release of thermonuclear energy.\\
To illustrate the mechanisms at work here, we consider the target design described in Ref. \cite{2024_Temporal_epjplus}. The implosion is simulated using the hydrocode MULTI-IFE\cite{2016_Ramis_cpc} with the focus on the deceleration phase. This design delivers 27 MJ of thermonuclear energy in 1D, for 355 kJ of absorbed laser energy and 300 µg of DT fuel. The burning rate is approximately 27 $\%$ and its distinctive feature is that it lies relatively far above its kinetic threshold.\\
The effective way of illustrating the onset of hot layer ignition is to plot the spatiotemporal evolution of the fusion power per unit mass during the end of implosion and combustion (see Fig. \ref{simul}). 
We can visualize the target imploding as the shock wave detaches from the hot spot and propagates through the layer in free fall. By zooming in on the portion preceding recompression during the expansion of the hot spot, we can see the ignition of the shocked zone, followed by the recompression of the fuel.\\
The ignition of this shocked shell must therefore satisfy the ignition criterion of equation (\ref{couche_critere}) somewhere in time. The figure \ref{ign_crit} compares the time evolution of the areal density of the shocked shell $\rho_{ss}\Delta e_{ss}$ as a function of its ion temperature over time, between stagnation and recompression extracted from simulation. On this same graph, the ignition criterion derived from equation (\ref{couche_critere}) is also plotted for different time values. The parameter $\eta_{e,ss}=2$ is determined by taking into account the bell-shaped radial temperature profile of the shocked shell during this time interval, which retains this shape throughout the $2$ to $3$ ps ignition duration. As $\eta_{rad}$ is a function of $\rho_{ss} \Delta e_{ss}$ and $T_{e,ss}$, Eq.(\ref{couche_critere}) is solved self-consistently with Monte Carlo evaluation of $\eta_{rad}$. We also suppose $T_{e,ss}=T_{i,ss}$.\\
The areal density of the shocked shell satisfies the ignition criterion by crossing the criterion at the time 16.2 ns. Once the ignition condition is met, it takes time (approximately 2 ps) for the shell pressure to overcome hot-spot pressure and compress the hot spot once again.\\
Surprisingly before its ignition, the shocked shell areal density decreases after reaching a maximum, while the thickness of the shocked shell is diminishing. This is due to a competition between losses and gains of mass in the shocked shell. Indeed, the propagation of the hot spot front within the shocked shell is driven by thermal conduction and the energy deposition of alpha particles at the edge of the hot spot. This then supplies the hot spot with fuel while removing mass from the shocked shell, as the compressed portion situated between the hot spot and the impacted shell advances into the latter. At the same time, the propagation of the shock, governed by the Rankine-Hugoniot equations, accretes mass into the shocked shell. There is therefore a competition between matter leaving the shocked region to go in the hot spot and matter entering through the shock front. This competition could therefore be directly induced by the heat flux flowing from the hot spot towards the density ridge separating the hot spot from the shocked shell. Indeed, simulations varying the limiting factor of the heat flux at the hot spot have shown that, all other things being equal, the lower the heat flux, the lower the kinetic threshold for combustion \cite {2019_temporal_epjd}. This supports the idea that heat flux is one of the key parameters for ignition of the hot layer and combustion. A reduction in heat transfer from the hot spot to the density ridge should lower the level of the ridge and increase the shocked-shell areal density, assuming that the divergent shock propagates at the same speed. In this way, the ignition criterion could then be met earlier in the implosion and at lower ionic temperatures.\\
Furthermore, when the simulation (with MULTI-IFE) is carried out by cutting off the energy transfer from the alpha particles in the shocked shell, the total thermonuclear energy released by the simulation is 18 MJ, compared with 27 MJ when the alpha particles re-transfer their energy. The total gain, defined as the ratio of fusion energy to laser energy, falls from 77 to 50. In this case, there is no recompression of the hot spot. Igniting this hot layer therefore appears to be a key factor in achieving the sufficiently high gains – in excess of 100 – required to consider energy production via inertial fusion\cite{2015_Atzeni_epl,2025_Ribeyre_AIPadvance}.\\
Finally, we performed a MULTI-IFE simulation, assuming local deposition of alpha particles (deposited at the same location where they are created), and we observe the same behavior: the hot layer ignites and subsequently recompresses the hot spot.\\
Unfortunately, unlike hot-spot ignition, hot layer ignition is particularly sensitive to preheating by hot electrons, as previously observed but unexplained in Ref.\cite{2021_temporal_epjd,2024_Temporal_epjplus}. Changes in areal density when electrons deposit their energy in the compressed shell prevent the ignition criterion from being met and, consequently, prevent combustion. However, the fact that the use of a shock ignition peak \cite{2024_temporal_jpp} succeeds in igniting a target in the presence of hot electrons – a target that does not burn without this peak – is also indicative of a change in the areal density of the hot layer, enabling the ignition criterion to be met.\\
Finally, just like hot-spot ignition, hot layer ignition is also sensitive to low-mode deformations and hydrodynamic instabilities\cite{2026_Trickey_pop}, which would therefore tend to prevent the ignition criterion for the shocked shell from being met.\\
To conclude, the process described in this letter differs from conventional burn propagation in that the dominant source is distributed across a finite-volume layer already subjected to a shock, rather than being confined to a narrow front propagating from the hot spot in imploding fuel. Ignition is volumetric and therefore differs from the commonly accepted picture of a reactive burning shock front.

\begin{acknowledgments}
This work has been done under the auspices of CEA-DAM.\\
Authors would like to thank L. Masse for fruitful discussions and R. Ramis and M. temporal for providing MULTI-IFE code.\\
The authors has no conflicts to disclose.\\
The authors acknowledge Pr. S. Fujioka for endorsing the draft on arXiv.\\
The data that support the findings of this study are available from the corresponding author upon reasonable request.
\end{acknowledgments}

\bibliography{biblio}

\end{document}